\documentclass[sigconf]{acmart}
\AtBeginDocument{%
  \providecommand\BibTeX{{%
    \normalfont B\kern-0.5em{\scshape i\kern-0.25em b}\kern-0.8em\TeX}}}

\setcopyright{iw3c2w3}
\copyrightyear{2020}
\acmYear{2020}
\acmDOI{10.1145/3366423.3380297}

\acmConference[WWW '20]{Proceedings of The Web Conference 2020}{April 20--24, 2020}{Taipei, Taiwan}
\acmBooktitle{Proceedings of The Web Conference 2020 (WWW '20), April 20--24, 2020, Taipei, Taiwan}
\acmPrice{}
\acmISBN{978-1-4503-7023-3/20/04}
\usepackage{multirow}
\usepackage{bm}
\usepackage{amsthm}
\usepackage{subcaption} % not in ACM TAPS whitelist
\usepackage[linesnumbered,ruled]{algorithm2e}

\theoremstyle{definition}
\newtheorem{definition}{Definition}[section]

\newcommand{\specialcell}[2][c]{%
  \begin{tabular}[#1]{@{}c@{}}#2\end{tabular}}

\begin{document}

\title{MAGNN: Metapath Aggregated Graph Neural Network for Heterogeneous Graph Embedding}

\author{Xinyu Fu}
\affiliation{%
  \institution{The Chinese University of Hong Kong}
  \city{Hong Kong}
  \country{China}}
\email{xyfu@cse.cuhk.edu.hk}

\author{Jiani Zhang}
\affiliation{%
  \institution{The Chinese University of Hong Kong}
  \city{Hong Kong}
  \country{China}}
\email{jnzhang@cse.cuhk.edu.hk}

\author{Ziqiao Meng}
\affiliation{%
  \institution{The Chinese University of Hong Kong}
  \city{Hong Kong}
  \country{China}}
\email{zqmeng@cse.cuhk.edu.hk}

\author{Irwin King}
\affiliation{%
  \institution{The Chinese University of Hong Kong}
  \city{Hong Kong}
  \country{China}}
\email{king@cse.cuhk.edu.hk}

\renewcommand{\shortauthors}{Fu et al.}

\begin{abstract}
A large number of real-world graphs or networks are inherently heterogeneous, involving a diversity of node types and relation types. Heterogeneous graph embedding is to embed rich structural and semantic information of a heterogeneous graph into low-dimensional node representations. Existing models usually define multiple metapaths in a heterogeneous graph to capture the composite relations and guide neighbor selection. However, these models either omit node content features, discard intermediate nodes along the metapath, or only consider one metapath. To address these three limitations, we propose a new model named \emph{Metapath Aggregated Graph Neural Network} (MAGNN) to boost the final performance. Specifically, MAGNN employs three major components, i.e., the node content transformation to encapsulate input node attributes, the intra-metapath aggregation to incorporate intermediate semantic nodes, and the inter-metapath aggregation to combine messages from multiple metapaths. Extensive experiments on three real-world heterogeneous graph datasets for node classification, node clustering, and link prediction show that MAGNN achieves more accurate prediction results than state-of-the-art baselines.
\end{abstract}





\maketitle

\section{Introduction}

Many real-world datasets are naturally represented in a graph data structure, where objects and the relationships among them are embodied by nodes and edges, respectively. Examples include social networks~\cite{DBLP:conf/kdd/WangC016, DBLP:conf/nips/HamiltonYL17}, physical systems~\cite{NIPS2016_6418, NIPS2017_7231}, traffic networks~\cite{li2018diffusion, DBLP:conf/uai/ZhangSXMKY18}, citation networks~\cite{NIPS2016_6212, DBLP:conf/iclr/KipfW17, DBLP:conf/nips/HamiltonYL17}, recommender systems~\cite{DBLP:journals/corr/BergKW17, DBLP:conf/ijcai/ZhangSZK19}, knowledge graphs~\cite{NIPS2013_5071, DBLP:conf/iclr/SunDNT19}, and so on.
The unique non-Euclidean nature of graphs renders them difficult to be modeled by traditional machine learning models.
For the neighborhood set of each node, there is no order or size limit. However, most statistical models assume an ordered and fixed-size input lying in the Euclidean space.
Therefore, it would be beneficial if nodes could be represented by meaningful low-dimensional vectors in the Euclidean space and then be taken as the input for other machine learning models.

Different graph embedding techniques have been proposed for the graph structure.
LINE~\cite{Tang:2015:LLI:2736277.2741093} generates node embeddings by exploiting the first-order and second-order proximity between nodes.
Random-walk-based methods including DeepWalk~\cite{Perozzi:2014:DOL:2623330.2623732}, node2vec~\cite{Grover:2016:NSF:2939672.2939754}, and TADW~\cite{Yang:2015:NRL:2832415.2832542} feed node sequences generated by random walks to a skip-gram model~\cite{DBLP:journals/corr/abs-1301-3781} to learn node embeddings.
With the rapid development of deep learning, graph neural networks (GNNs) have been proposed, which learn the graph representations using specially designed neural layers.
Spectral-based GNNs, including ChebNet~\cite{Defferrard:2016:CNN:3157382.3157527} and GCN~\cite{DBLP:conf/iclr/KipfW17}, perform graph convolution operations in the Fourier domain of an entire graph.
Recent spatial-based GNNs, including GraphSAGE~\cite{DBLP:conf/nips/HamiltonYL17}, GAT~\cite{DBLP:conf/iclr/VelickovicCCRLB18}, and many other variants~\cite{DBLP:journals/corr/LiTBZ15, DBLP:conf/uai/ZhangSXMKY18, DBLP:conf/ijcai/ZhangSZK19}, address the issues around scalability and generalization ability of the spectral-based models by performing graph convolution operations directly in the graph domain. An increasing number of researchers have paid attention to this promising area.

Although GNNs have achieved state-of-the-art results in many tasks, most GNN-based models assume that the input is a homogeneous graph with only one node type and one edge type.
Most real-world graphs consist of various types of nodes and edges associated with attributes in different feature spaces. For example, a co-authorship network contains at least two types of nodes, namely authors and papers.
Author attributes may include affiliations, citations, and research fields. Paper attributes may consist of keywords, venue, year, and so on.
We refer to graphs of this kind as \emph{heterogeneous information networks} (HINs) or \emph{heterogeneous graphs}.
The heterogeneity in both graph structure and node content makes it challenging for GNNs to encode their rich and diverse information into a low-dimensional vector space.

Most existing heterogeneous graph embedding methods are based on the idea of metapaths.
A \emph{metapath} is an ordered sequence of node types and edge types defined on the network schema, which describes a composite relation between the nodes types involved.
For example, in a scholar network with authors, papers, and venues, \textit{Author-Paper-Author} (APA) and \textit{Author-Paper-Venue-Paper-Author} (APVPA) are metapaths describing two different relations among authors. The APA metapath associates two co-authors, while the APVPA metapath associates two authors who published papers in the same venue. Therefore, we can view a metapath as high-order proximity between two nodes.
Because traditional GNNs treat all nodes equally, they are unable to model the complex structural and semantic information in heterogeneous graphs.

Although these metapath-based embedding methods outperform traditional network embedding methods on various tasks, such as node classification and link prediction, they still suffer from at least one of the following limitations.
(1) The model does not leverage node content features, so it rarely performs well on heterogeneous graphs with rich node content features (e.g., metapath2vec~\cite{Dong:2017:MSR:3097983.3098036}, ESim~\cite{DBLP:journals/corr/ShangQLKHP16}, HIN2vec~\cite{Fu:2017:HEM:3132847.3132953}, and HERec~\cite{8355676}).
(2) The model discards all intermediate nodes along the metapath by only considering two end nodes, which results in information loss (e.g., HERec~\cite{8355676} and HAN~\cite{Wang:2019:HGA:3308558.3313562}).
% For instance, movies acted by Sylvester Stallone would probably fall in the same action genre, while movies acted by Leonardo DiCaprio would have more diverse genres. Therefore, right after the node content transformation, we design the intra-metapath aggregation to learn the structural and semantic information embedded in the metapath-based neighbors and the context in between.
(3) The model relies on a single metapath to embed the heterogeneous graph. Hence, the model requires a manual metapath selection process and loses aspects of information from other metapaths, leading to suboptimal performance (e.g., metapath2vec~\cite{Dong:2017:MSR:3097983.3098036}).

To address these limitations, we propose a novel \emph{Metapath Aggregated Graph Neural Network} (MAGNN) for heterogeneous graph embedding.
MAGNN addresses all the issues described above by applying node content transformation, intra-metapath aggregation, and inter-metapath aggregation to generate node embeddings.
Specifically, MAGNN first applies type-specific linear transformations to project heterogeneous node attributes, with possibly unequal dimensions for different node types, to the same latent vector space.
Next, MAGNN applies intra-metapath aggregation with the attention mechanism~\cite{DBLP:conf/iclr/VelickovicCCRLB18} for every metapath.
During this intra-metapath aggregation, each target node extracts and combines information from the metapath instances connecting the node with its metapath-based neighbors.
In this way, MAGNN captures the structural and semantic information of heterogeneous graphs from both neighbor nodes and the metapath context in between.
Following intra-metapath aggregation, MAGNN further conducts inter-metapath aggregation using the attention mechanism to fuse latent vectors obtained from multiple metapaths into final node embeddings.
By integrating multiple metapaths, our model can learn the comprehensive semantics ingrained in the heterogeneous graph.

In summary, this work makes several major contributions:
\begin{enumerate}
    \item We propose a novel metapath aggregated graph neural network for heterogeneous graph embedding.
    \item We design several candidate encoder functions for distilling information from metapath instances, including one based on the idea of relational rotation in complex space~\cite{DBLP:conf/iclr/SunDNT19}.
    \item We conduct extensive experiments on the IMDb and the DBLP datasets for node classification and node clustering, as well as on the Last.fm dataset for link prediction to evaluate the performance of our proposed model. Experiments on all of these datasets and tasks show that the node embeddings learned by MAGNN are consistently better than those generated by other state-of-the-art baselines.
\end{enumerate}

\section{Preliminary}
\label{sec:preliminary}

\begin{figure*}[t]
    \centering
\begin{tabular}[b]{@{}c@{}}
    \begin{subfigure}[b]{0.26\textwidth}
        \centering
        \includegraphics[width=\textwidth]{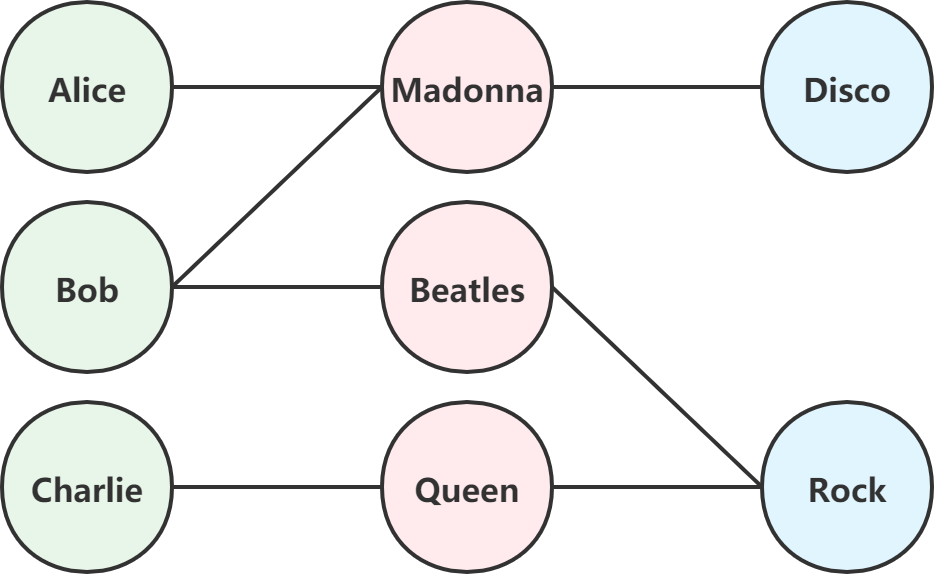}
        \caption{Heterogeneous Graph}
        \label{fig:het-graph}
    \end{subfigure} \\
    \begin{subfigure}[b]{0.36\textwidth}
        \centering
        \includegraphics[width=\textwidth]{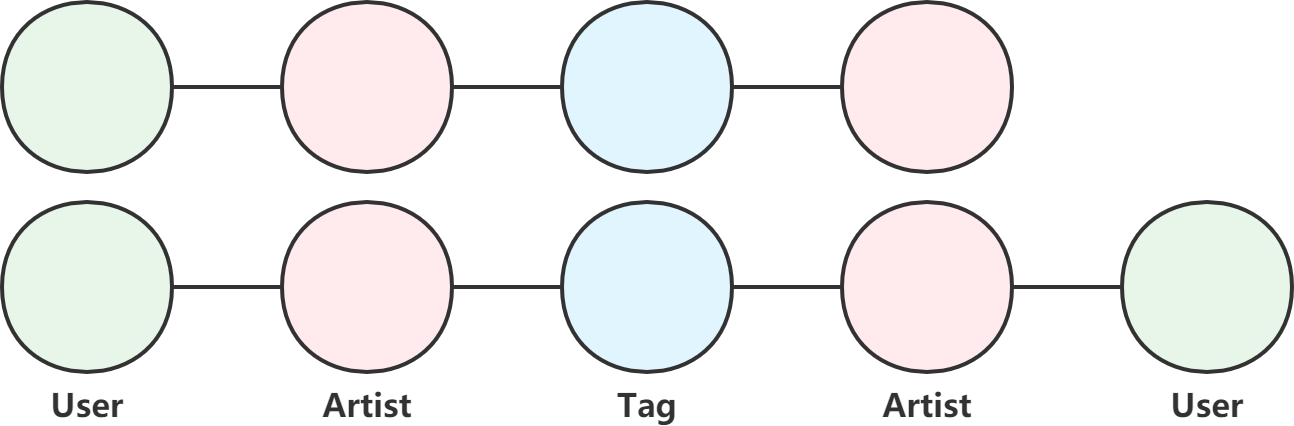}
        \caption{Metapaths}
        \label{fig:metapath}
    \end{subfigure}
\end{tabular}
\hfill
\begin{subfigure}[b]{0.36\textwidth}
    \centering
    \includegraphics[width=\textwidth]{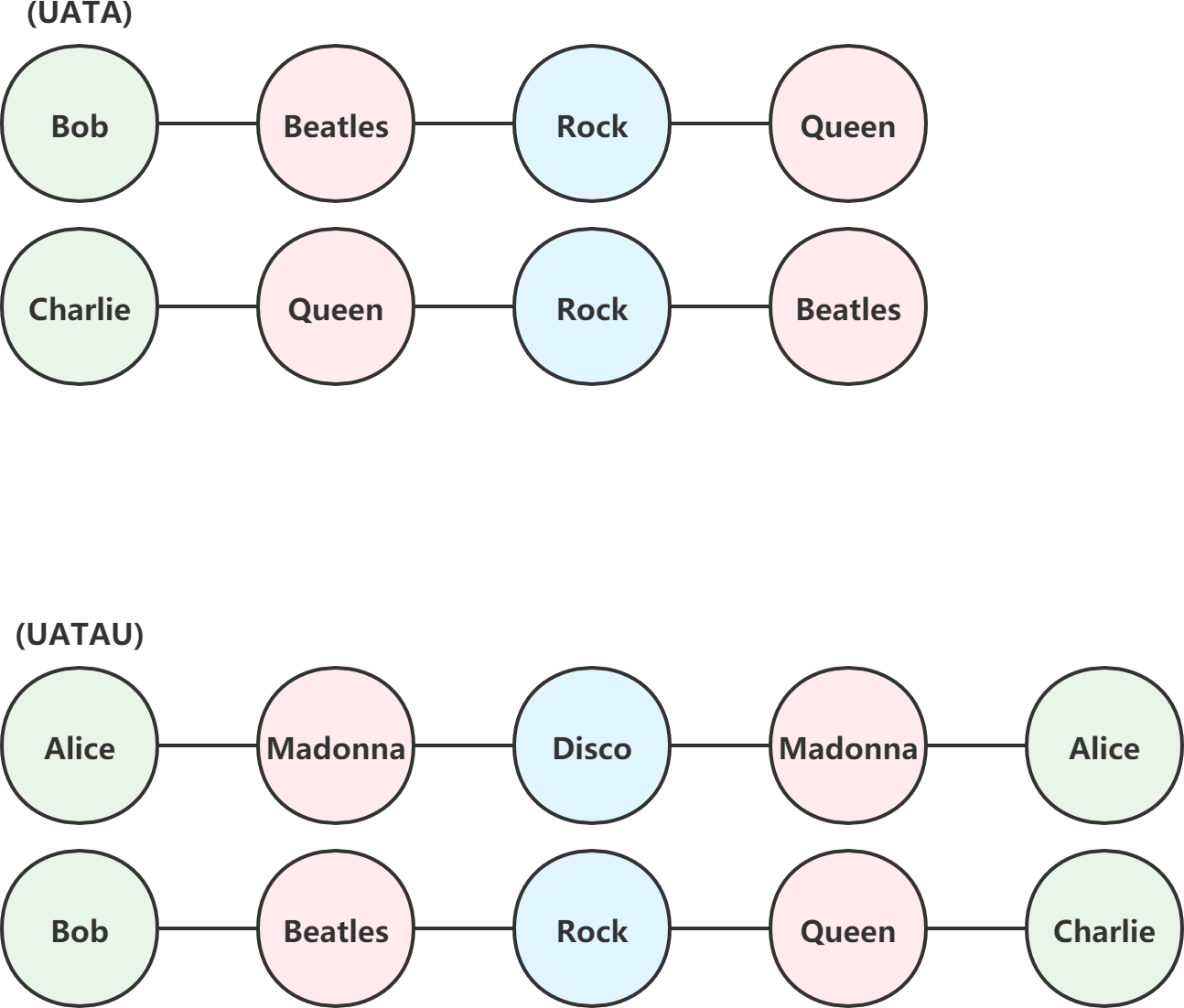}
    \caption{Metapath Instances}
    \label{fig:metapath-instance}
\end{subfigure}
\hfill
\begin{subfigure}[b]{0.19\textwidth}
    \centering
    \includegraphics[width=\textwidth]{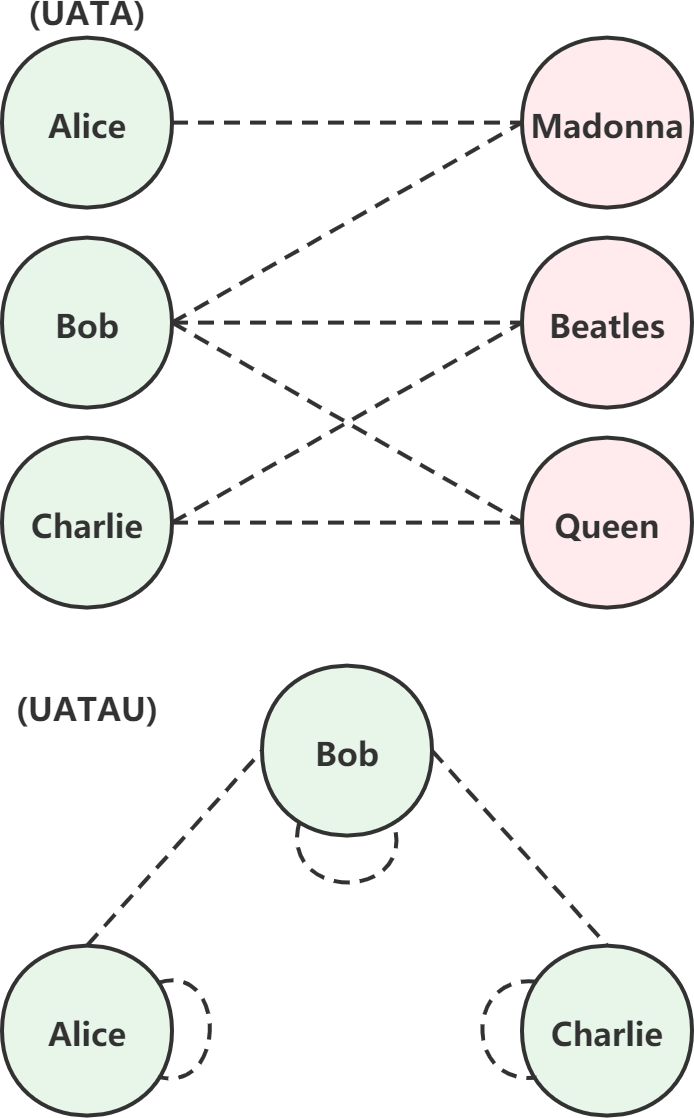}
    \caption{Metapath-based Graphs}
    \label{fig:metapath-graph}
\end{subfigure}
    \caption{An illustration of the terms defined in Section~\ref{sec:preliminary}. (a) An example heterogeneous graph with three types of nodes (i.e., users, artists, and tags). (b) The User-Artist-Tag-Artist (UATA) metapath and the User-Artist-Tag-Artist-User (UATAU) metapath. (c) Example metapath instances of the UATA and UATAU metapaths, respectively. (d) The metapath-based graphs for the UATA and UATAU metapaths, respectively.}
    \label{fig:illustration}
\end{figure*}

In this section, we give formal definitions of some important terminologies related to heterogeneous graphs.
Graphical illustrations are provided in Figure~\ref{fig:illustration}.
Besides, Table~\ref{tab:notation} summarizes frequently used notations in this paper for quick reference.

\begin{table}[!t]
    \caption{Notations used in this paper.}
    \label{tab:notation}
    %\centering
    \begin{tabular}{l|l}
        \toprule
        \textbf{Notations} & \textbf{Definitions}\\
        $\mathbb{R}^{n}$ & $n$-dimensional Euclidean space\\
        $a$, $\mathbf{a}$, $\mathbf{A}$ & Scalar, vector, matrix\\
        $\mathbf{A}^{\intercal}$ & Matrix/vector transpose\\
        $\mathcal{V}$ & The set of nodes in a graph\\
        $\mathcal{E}$ & The set of edges in a graph\\
        $\mathcal{G}$ & A graph $\mathcal{G}=\left(\mathcal{V},\mathcal{E}\right)$\\
        $v$ & A node $v\in\mathcal{V}$\\
%        $e_{ij}$ & An edge $e_{ij}=\left(v_i,v_j\right)\in\mathcal{E}$\\
%        $D_{v}$ & The degree of node $v$\\
%        $N$ & The number of nodes in a graph, $N=|\mathcal{V}|$\\
%        $M$ & The number of edges in a graph, $M=|\mathcal{E}|$\\
%        $\mathbf{A} \in \mathbb{R}^{N \times N}$ & The adjacency matrix of a graph\\
        $P$ & A metapath\\
        $P\left(v,u\right)$ & A metapath instance connecting node $v$ and $u$\\
        $\mathcal{N}_v$ & The set of neighbors of node $v$\\
        $\mathcal{N}_v^{P}$ & The set of metapath-$P$-based neighbors of node $v$\\
%        $\mathcal{E}_v$ & The set of edges connected to node $v$\\
%        $N\left(v\right)$ & The set of neighbors of node $v$\\
%        $E\left(v\right)$ & The set of edges connected to node $v$\\
        $\mathbf{x}_v$ & Raw (content) feature vector of node $v$\\
%        $\mathbf{X}$ & The raw features of all nodes in a graph\\
        $\mathbf{h}_v$ & Hidden state (embedding) of node $v$\\
%        $\mathbf{h}_v^l$ & The hidden state (embedding) of node $v$ at layer $l$\\
        $\mathbf{W}$ & Weight matrix\\
        $\alpha, \beta$ & Normalized attention weight\\
%        $\mathbf{h}_{N\left(v\right)}^{t}$ & The aggregated neighborhood vector of node $v$ at layer $t$\\
%        $\mathbf{H}$ & The hidden states of all nodes in a graph\\
%        $\mathbf{H}^{t}$ & The hidden states of all nodes in a graph at layer t\\
%        $\bm{\Theta}$, $\mathbf{W}$ & The learnable parameters/weights of a model\\
        $\sigma(\cdot)$ & Activation function\\
        $\odot$ & Element-wise multiplication\\
        $|\cdot|$ & The cardinality of a set\\
        $\|$ & Vector concatenation\\
        \bottomrule
    \end{tabular}
\end{table}

\begin{definition}{\textbf{Heterogeneous Graph.}}
A heterogeneous graph is defined as a graph $\mathcal{G}=\left(\mathcal{V},\mathcal{E}\right)$ associated with a node type mapping function $\phi : \mathcal{V} \rightarrow \mathcal{A}$ and an edge type mapping function $\psi : \mathcal{E} \rightarrow \mathcal{R}$.
$\mathcal{A}$ and $\mathcal{R}$ denote the predefined sets of node types and edge types, respectively, with $|\mathcal{A}|+|\mathcal{R}|>2$.
\end{definition}

%For example, Figure~\ref{fig:het-graph} is a heterogeneous graph with three node types (i.e., users, artists, and tags) and two edge types (i.e., user-artist edge and artist-tag edge).

\begin{definition}{\textbf{Metapath.}}
A metapath $P$ is defined as a path in the form of $A_{1} \stackrel{R_{1}}{\longrightarrow} A_{2} \stackrel{R_{2}}{\longrightarrow} \cdots \stackrel{R_{l}}{\longrightarrow} A_{l+1}$ (abbreviated as $A_{1} A_{2} \cdots A_{l+1}$), which describes a composite relation $R=R_{1} \circ R_{2} \circ \cdots \circ R_{l}$ between node types $A_1$ and $A_{l+1}$, where $\circ$ denotes the composition operator on relations.
\end{definition}

%For example, Figure~\ref{fig:metapath} illustrates the \textit{User-Artist-Tag-Artist} (UATA) and \textit{User-Artist-Tag-Artist-User} (UATAU) metapaths in the heterogeneous graph given in Figure~\ref{fig:het-graph}. The UATA metapath associates a user and an artist %TODO. The UATAU metapath associates two users listening to artists with the same tag.

\begin{definition}{\textbf{Metapath Instance.}}
Given a metapath $P$ of a heterogeneous graph, a metapath instance $p$ of $P$ is defined as a node sequence in the graph following the schema defined by $P$.
\end{definition}

\begin{definition}{\textbf{Metapath-based Neighbor.}}
Given a metapath $P$ of a heterogeneous graph, the metapath-based neighbors $\mathcal{N}^P_v$ of a node $v$ is defined as the set of nodes that connect with node $v$ via metapath instances of $P$. A neighbor connected by two different metapath instances is regarded as two different nodes in $\mathcal{N}^P_v$. Note that $\mathcal{N}^P_v$ includes $v$ itself if $P$ is symmetric.
\end{definition}

For example, considering the metapath UATA in Figure~\ref{fig:illustration}, artist \textit{Queen} is a metapath-based neighbor of user \textit{Bob}. These two nodes are connected via the metapath instance \textit{Bob-Beatles-Rock-Queen}. Moreover, we may refer to \textit{Beatles} and \textit{Rock} as the intermediate nodes along this metapath instance.

\begin{definition}{\textbf{Metapath-based Graph.}}
Given a metapath $P$ of a heterogeneous graph $\mathcal{G}$, the metapath-based graph $\mathcal{G}^P$ is a graph constructed by all the metapath-$P$-based neighbor pairs in graph $\mathcal{G}$. Note that $\mathcal{G}^P$ is homogeneous if $P$ is symmetric.
\end{definition}

\begin{definition}{\textbf{Heterogeneous Graph Embedding.}}
Given a heterogeneous graph $\mathcal{G}=\left(\mathcal{V},\mathcal{E}\right)$, with node attribute matrices $\mathbf{X}_{A_i} \in \mathbb{R}^{|\mathcal{V}_{A_i}| \times d_{A_i}}$ for node types $A_i \in \mathcal{A}$, heterogeneous graph embedding is the task to learn the $d$-dimensional node representations $\mathbf{h}_{v} \in \mathbb{R}^{d}$ for all $v \in \mathcal{V}$ with $d \ll |\mathcal{V}|$ that are able to capture rich structural and semantic information involved in $\mathcal{G}$.
\end{definition}

\section{Related Work}

In this section, we review studies on graph representation learning that are related to our model. They are organized into two subsections: Section~\ref{sec:GNN_survey} summarizes research efforts on GNNs for general graph embedding, while Section~\ref{sec:heter_survey} introduces graph embedding methods designed for heterogeneous graphs.

\subsection{Graph Neural Networks}
\label{sec:GNN_survey}

The goal of a GNN is to learn a low-dimensional vector representation $\mathbf{h}_v$ for every node $v$, which can be used for many downstream tasks, e.g., node classification, node clustering, and link prediction.
The rationale behind this is that each node is naturally defined by its own features and its neighborhood.
Following this idea and based on graph signal processing, spectral-based GNNs were first developed to perform graph convolution in the Fourier domain of a graph.
ChebNet~\cite{Defferrard:2016:CNN:3157382.3157527} utilizes Chebyshev polynomials to filter graph signals (node features) in the graph Fourier domain. Another influential model of this kind is GCN~\cite{DBLP:conf/iclr/KipfW17}, which constrains and simplifies the parameters of ChebNet to alleviate the overfitting problem and improve the performance.
However, spectral-based GNNs suffer from poor scalability and generalization ability, because they require the entire graph as input for every layer, and their learned filters depend on the eigenbasis of the graph Laplacian, which is closely related to the specific graph structure.

Spatial-based GNNs have been proposed to address these two limitations.
GNNs of this kind define convolutions directly in the graph domain by aggregating feature information from neighbors for each node, thus imitating the convolution operations of convolutional neural networks for image data.
GraphSAGE~\cite{DBLP:conf/nips/HamiltonYL17}, the seminal spatial-based GNN framework, is founded upon the general notion of aggregator functions for efficient generation of node embeddings.
The aggregator function samples, extracts, and transforms a target node's local neighborhood, and thus facilitates parallel training and generalization to unseen nodes or graphs.
Many other spatial-based GNN variants have been proposed based on this idea.
Inspired by the Transformer~\cite{DBLP:conf/nips/VaswaniSPUJGKP17}, GAT~\cite{DBLP:conf/iclr/VelickovicCCRLB18} incorporates the attention mechanism into the aggregator function to take into account the relative importance of each neighbor's information from the target node's perspective.
GGNN~\cite{DBLP:journals/corr/LiTBZ15} adds a gated recurrent unit (GRU)~\cite{DBLP:journals/corr/ChoMGBSB14} to the aggregator function by treating the aggregated neighborhood information as the input to the GRU of the current time step.
GaAN~\cite{DBLP:conf/uai/ZhangSXMKY18} combines GRU with the gated multi-head attention mechanism for dealing with spatiotemporal graphs.
STAR-GCN~\cite{DBLP:conf/ijcai/ZhangSZK19} stacks multiple GCN encoder-decoders to boost the rating prediction performance.

All of the GNNs mentioned above are either built for homogeneous graphs, or designed for graphs with a special structure, as in user-item recommender systems. Because most existing GNNs operate on features of nodes in the same shared embedding space, they cannot be naturally adapted to heterogeneous graphs with node features lying in different spaces.

\subsection{Heterogeneous Graph Embedding}
\label{sec:heter_survey}

Heterogeneous graph embedding aims to project nodes in a heterogeneous graph into a low-dimensional vector space.
This challenging topic has been addressed by a number of studies.
For example, metapath2vec~\cite{Dong:2017:MSR:3097983.3098036} generates random walks guided by a single metapath, which are then fed to a skip-gram model~\cite{DBLP:journals/corr/abs-1301-3781} to generate node embeddings.
Given user-defined metapaths, ESim~\cite{DBLP:journals/corr/ShangQLKHP16} generates node embeddings by learning from sampled positive and negative metapath instances.
HIN2vec~\cite{Fu:2017:HEM:3132847.3132953} carries out multiple prediction training tasks to learn representations of nodes and metapaths of a heterogeneous graph.
Given a metapath, HERec~\cite{8355676} converts a heterogeneous graph into a homogeneous graph based on metapath-based neighbors and applies the DeepWalk model to learn the node embeddings of the target type.
Like HERec, HAN~\cite{Wang:2019:HGA:3308558.3313562} converts a heterogeneous graph into multiple metapath-based homogeneous graphs in a similar way, but uses a graph attention network architecture to aggregate information from the neighbors and leverages the attention mechanism to combine various metapaths.
Another model, PME~\cite{Chen:2018:PPM:3219819.3219986}, learns node embeddings by projecting them into the corresponding relation spaces and optimizing the proximity between the projected nodes.

However, all of the heterogeneous graph embedding methods introduced above have the limitations of either ignoring node content features, discarding all intermediate nodes along the metapath, or utilizing only a single metapath.
Although they might have improved upon the performance of homogeneous graph embedding methods for some heterogeneous graph datasets, there is still room for improvement by exploiting more comprehensively the information embedded in heterogeneous graphs.

\begin{figure*}[ht!]
     \centering
     \hfill
     \begin{subfigure}[b]{0.41\textwidth}
         \centering
         \includegraphics[width=\textwidth]{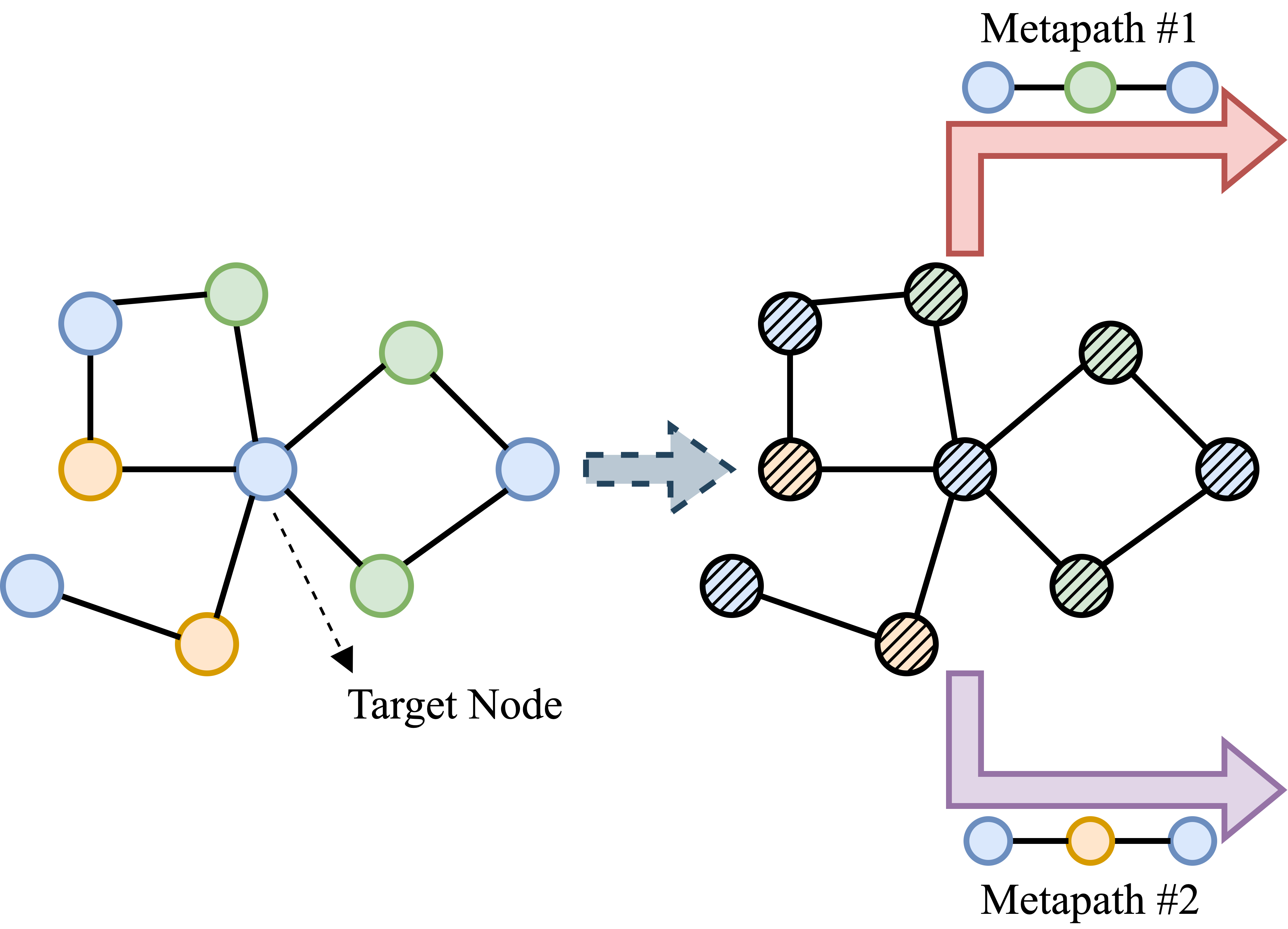}
         \caption{Node Content Transformation}
     \end{subfigure}
     %\hfill
     \begin{subfigure}[b]{0.3\textwidth}
         \centering
         \includegraphics[width=\textwidth]{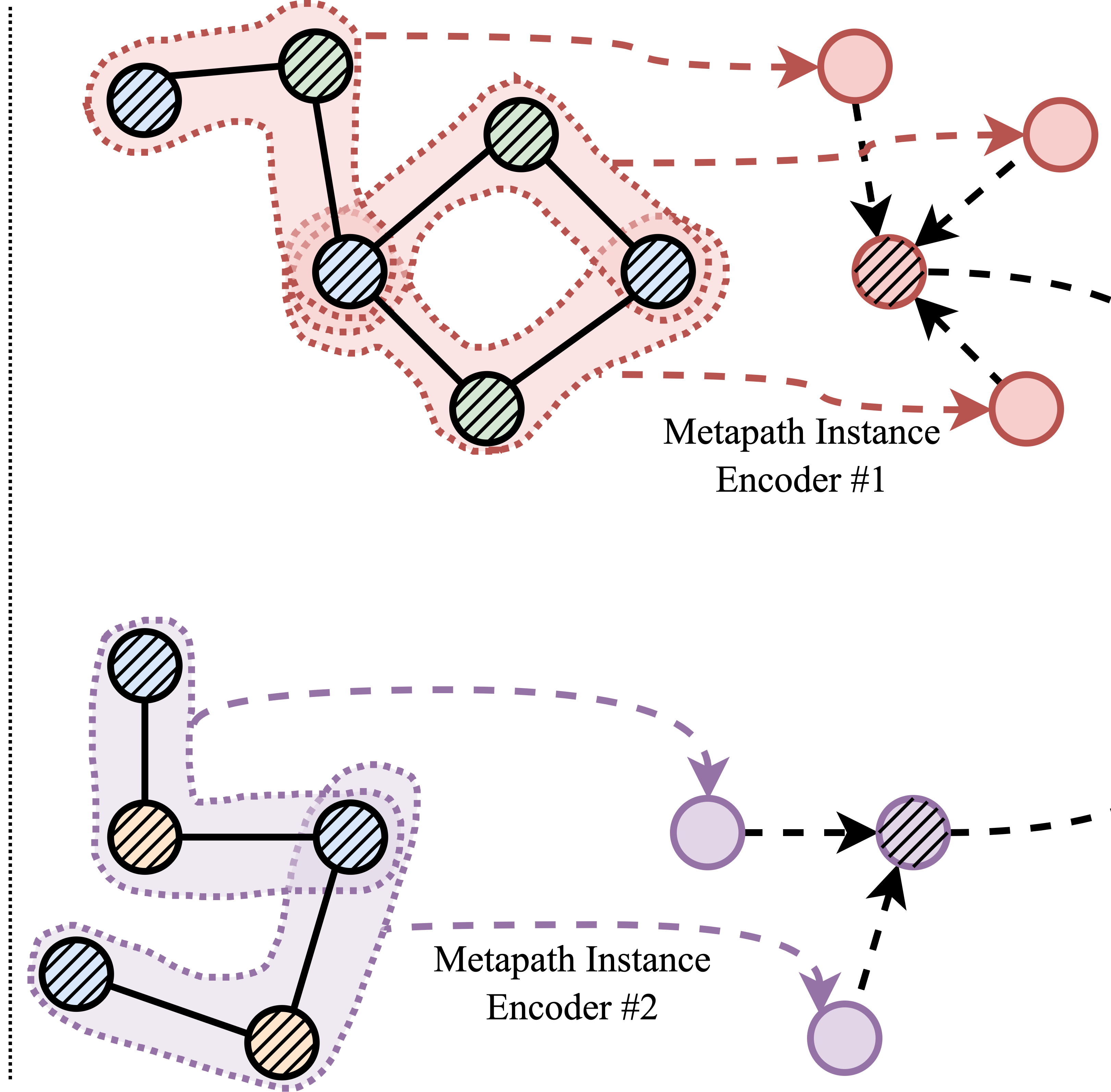}
         \caption{Intra-metapath Aggregation}
     \end{subfigure}
     %\hfill
     \begin{subfigure}[b]{0.277\textwidth}
         \centering
         \includegraphics[width=\textwidth]{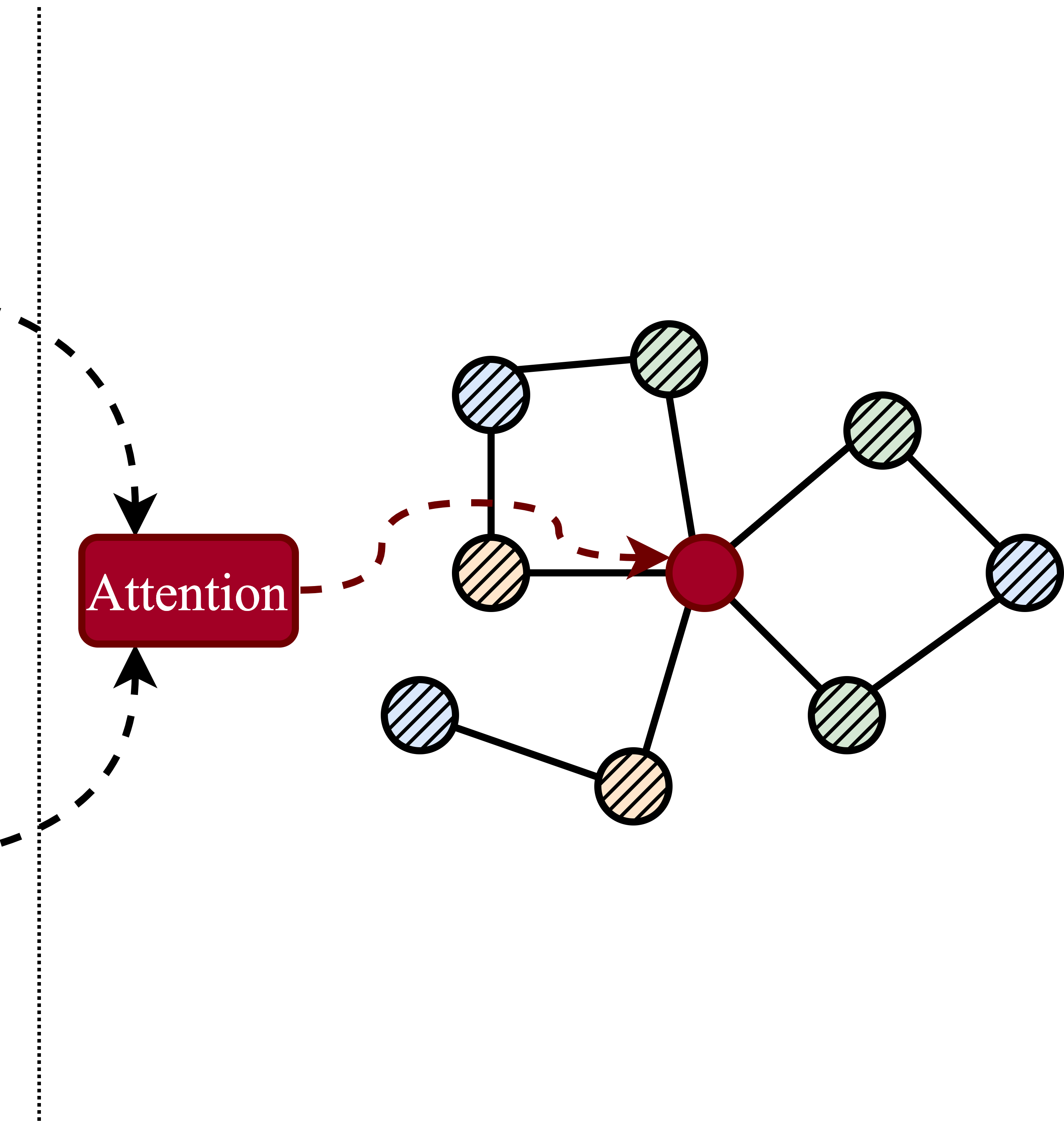}
         \caption{Inter-metapath Aggregation}
     \end{subfigure}
     \hfill
      \caption{The overall architecture of MAGNN (path instances that start and end with the target node are omitted for clarity).}
      \label{fig:MAGNN}
\end{figure*}

\section{Methodology}
\label{sec:methodology}

% a boost of performance
% exploit: utilize, leverage, use
% exhibit: capture, encode, learn
% adhere to: follow, comply with
% extractor or encoder

In this section, we describe a new metapath aggregated graph neural network (MAGNN) for heterogeneous graph embedding.
MAGNN is constructed by three major components: node content transformation, intra-metapath aggregation, and inter-metapath aggregation.
Figure~\ref{fig:MAGNN} illustrates the embedding generation of a single node. The overall forward propagation process is shown in Algorithm~\ref{algo:MAGNN}.

%The key rationale behind our approach is that aggregating projected features solely from metapath-based neighbors would cause information loss due to dropping the paths connecting two nodes.
%To tackle this problem, we define a particular metapath instance encoder function to incorporate the information embedded in the metapath context between the target node and its metapath-based neighbors, during the node-level aggregation.
%After that, we can fuse metapath-specific node vectors into final node embeddings with more representation power.

\subsection{Node Content Transformation}
\label{sec:node_type_specific}

For a heterogeneous graph associated with node attributes, different node types may have unequal dimensions of feature vectors. Even if they happen to be the same dimension, they may lie in different feature spaces. For example, $n_{1}$-dimensional bag-of-words vectors of texts and $n_{2}$-dimensional intensity histogram vectors of images cannot directly operate together even if $n_{1}=n_{2}$. Feature vectors of different dimensions are troublesome when we process them in a unified framework. Therefore, we need to project different types of node features into the same latent vector space before all else.

So before feeding node vectors into MAGNN, we apply a type-specific linear transformation for each type of nodes by projecting feature vectors into the same latent factor space.
For a node $v \in \mathcal{V}_A$ of type $A \in \mathcal{A}$, we have
\begin{equation}
    \mathbf{h}_{v}^{\prime}=\mathbf{W}_{A} \cdot \mathbf{x}_{v}^A,
\end{equation}
where $\mathbf{x}_{v} \in \mathbb{R}^{d_{A}}$ is the original feature vector, and $\mathbf{h}_{v}^{\prime} \in \mathbb{R}^{d^{\prime}}$  is the projected latent vector of node $v$. $\mathbf{W}_{A} \in \mathbb{R}^{d^{\prime} \times d_{A}}$ is the parametric weight matrix for type $A$'s nodes.

The node content transformation addresses the heterogeneity of a graph that originates from the node content features. After applying this operation, all nodes' projected features share the same dimension, which facilitates the aggregation process of the next model component.

\subsection{Intra-metapath Aggregation}
\label{sec:node_level}

Given a metapath $P$, the intra-metapath aggregation layer learns the structural and semantic information embedded in the target node, the metapath-based neighbors, and the context in between, by encoding the metapath instances of $P$.
Let $P(v,u)$ be a metapath instance connecting the \emph{target node} $v$ and the \emph{metapath-based neighbor} $u \in \mathcal{N}_{v}^{P}$, we further define the \emph{intermediate nodes} of $P(v,u)$ as $\{m^{P(v, u)}\} = P(v, u) \setminus \{u, v\}$.
Intra-metapath aggregation employs a special \emph{metapath instance encoder} to transform all the node features along a metapath instance into a single vector,
\begin{equation}
        \mathbf{h}_{P\left(v,u\right)} = f_{\theta}\left(P(v, u)\right) = f_{\theta}\left(\mathbf{h}_{v}^{\prime}, \mathbf{h}_{u}^{\prime}, \left\{\mathbf{h}_{t}^{\prime}, \forall t \in \{m^{P(v, u)}\} \right\}\right),
    \end{equation}
where $\mathbf{h}_{P\left(v,u\right)} \in \mathbb{R}^{d^{\prime}}$ has a dimension of $d'$.
For simplicity, here we use $P\left(v,u\right)$ to represent a single instance, although there might be multiple instances connecting the two nodes. Section~\ref{sec:metapath-instance-encoders} introduces several choices of a qualified metapath instance encoder.

After encoding the metapath instances into vector representations, we adopt a graph attention layer~\cite{DBLP:conf/iclr/VelickovicCCRLB18} to weighted sum the metapath instances of $P$ related to target node $v$.
The key idea is that different metapath instances would contribute to the target node's representation in different degrees. We can model this by learning a normalized importance weight $\alpha_{v u}^{P}$ for each metapath instance and then weighted summing all instances:
\begin{equation}
\label{eq:attn_1}
\begin{aligned}
e_{v u}^{P}&=\operatorname{LeakyReLU}\left(\mathbf{a}_{P}^{\intercal} \cdot\left[\mathbf{h}_{v}^{\prime} \| \mathbf{h}_{P\left(v,u\right)}\right]\right), \\
\alpha_{v u}^{P} &= \frac{\exp \left(e_{v u}^{P}\right)}{\sum_{s \in \mathcal{N}_{v}^{P}} \exp \left(e_{v s}^{P}\right)},\\
\mathbf{h}_{v}^{P} &= \sigma\left(\sum_{u \in \mathcal{N}_{v}^{P}} \alpha_{v u}^{P} \cdot \mathbf{h}_{P\left(v,u\right)}\right).
\end{aligned}
\end{equation}
Here $\mathbf{a}_{P} \in \mathbb{R}^{2d^{\prime}}$ is the parameterized attention vector for metapath $P$, and $\|$ denotes the vector concatenation operator. $e_{v u}^{P}$ indicates the importance of metapath instance $P\left(v,u\right)$ to node $v$, which is then normalized across all choices of $u \in \mathcal{N}_{v}^{P}$ using the softmax function. Once the normalized importance weight $\alpha_{v u}^{P}$ is obtained for all $u \in \mathcal{N}_{v}^{P}$, they are used to compute a weighted combination of the representations of the metapath instances about node $v$. Finally, the output goes through an activation function $\sigma(\cdot)$.

This attention mechanism can also be extended to multiple heads, which helps to stabilize the learning process and reduce the high variance introduced by the heterogeneity of graphs. That is, we execute $K$ independent attention mechanisms, and then concatenate their outputs, resulting in the following formulation:
\begin{equation}
    \mathbf{h}_{v}^{P} = \overset{K}{\underset{k=1}{\|}}\sigma\left(\sum_{u \in \mathcal{N}_{v}^{P}} \left[\alpha_{v u}^{P}\right]_{k} \cdot \mathbf{h}_{P\left(v,u\right)}\right),
\end{equation}
where $\left[\alpha_{v u}^{P}\right]_{k}$ is the normalized importance of metapath instance $P\left(v,u\right)$ to node $v$ at the $k$-th attention head.

To sum up, given the projected feature vectors $\mathbf{h}_{u}^{\prime} \in \mathbb{R}^{d^{\prime}} \forall u \in \mathcal{V}$ and the set of metapaths $\mathcal{P}_{A}=\left\{P_{1}, P_{2}, \ldots, P_{M}\right\}$ which start or end with node type $A \in \mathcal{A}$, the intra-metapath aggregation of MAGNN generates $M$ metapath-specific vector representations of the target node $v \in \mathcal{V}_{A}$, denoted as $\left\{\mathbf{h}_{v}^{P_{1}}, \mathbf{h}_{v}^{P_{2}}, \ldots, \mathbf{h}_{v}^{P_{M}}\right\}$. Each $\mathbf{h}_{v}^{P_{i}} \in \mathbb{R}^{d^{\prime}}$ (assuming $K=1$) can be interpreted as a summarization of the $P_{i}$-metapath instances about node $v$, exhibiting one aspect of semantic information contained in node $v$.

%%%%%%%%%%%%%% Jenny Mark: revision stopped here

\subsection{Inter-metapath Aggregation}
\label{sec:meta_path_level}

%By and large, each metapath of a heterogeneous graph describes one aspect of semantic information. A naive approach adopted by some embedding methods like metapath2vec~\cite{Dong:2017:MSR:3097983.3098036} is to select a single metapath empirically, which abandons many other metapaths that may be beneficial to the final node embeddings. The information of a node in a heterogeneous graph should be a combination of semantic information revealed by multiple metapaths. Based on this idea, MAGNN employs the inter-metapath aggregation to combine node representations generated from different metapaths into one meaningful node embedding.

After aggregating the node and edge data within each metapath, we need to combine the semantic information revealed by all metapaths using an inter-metapath aggregation layer. Now for a node type $A$, we have $|\mathcal{V}_A|$ sets of latent vectors: $\left\{\mathbf{h}_{v}^{P_{1}}, \mathbf{h}_{v}^{P_{2}}, \ldots, \mathbf{h}_{v}^{P_{M}}\right\}$ for $v\in \mathcal{V}_A$, where $M$ is the number of metapaths for type $A$. One straightforward inter-metapath aggregation approach is to take the element-wise mean of these node vectors. We extend this approach by exploiting the attention mechanism to assign different weights to different metapaths. This operation is reasonable because metapaths are not equally important in a heterogeneous graph.

First, we summarize each metapath $P_{i}\in \mathcal{P}_{A}$ by averaging the transformed metapath-specific node vectors for all nodes $v \in \mathcal{V}_A$,
\begin{equation}
    \mathbf{s}_{P_{i}}=\frac{1}{|\mathcal{V}_{A}|} \sum_{v \in \mathcal{V}_{A}} \tanh \left(\mathbf{M}_{A} \cdot \mathbf{h}_{v}^{P_{i}}+\mathbf{b}_{A}\right),
\end{equation}
where $\mathbf{M}_{A} \in \mathbb{R}^{d_{m} \times d^{\prime}}$ and $\mathbf{b}_{A} \in \mathbb{R}^{d_{m}}$ are learnable parameters.

Then we use the attention mechanism to fuse the metapath-specific node vectors of $v$ as follows:
\begin{equation}
\begin{aligned}
    & e_{P_{i}}=\mathbf{q}_{A}^{\intercal} \cdot \mathbf{s}_{P_{i}},\\
    & \beta_{P_{i}}=\frac{\exp \left(e_{P_{i}}\right)}{\sum_{P \in \mathcal{P}_{A}} \exp \left(e_{P}\right)}, \\
    & \mathbf{h}_{v}^{\mathcal{P}_{A}} = \sum_{P \in \mathcal{P}_{A}} \beta_{P} \cdot \mathbf{h}_{v}^{P},
\end{aligned}
\end{equation}
where $\mathbf{q}_{A} \in \mathbb{R}^{d_{m}}$ is the parameterized attention vector for node type $A$. $\beta_{P_{i}}$ can be interpreted as the relative importance of metapath $P_i$ to type $A$'s nodes. Once $\beta_{P_{i}}$ is computed for each $P_{i} \in \mathcal{P}_{A}$, we weighted sum all the metapath-specific node vectors of $v$.

At last, MAGNN employs an additional linear transformation with a nonlinear function to project the node embeddings to the vector space with the desired output dimension:
\begin{equation}
\label{eq:output_proj}
    \mathbf{h}_{v} = \sigma\left(\mathbf{W}_{o} \cdot \mathbf{h}_{v}^{\mathcal{P}_{A}}\right),
\end{equation}
where $\sigma(\cdot)$ is an activation function, and $\mathbf{W}_{o} \in \mathbb{R}^{d_o \times d^{\prime}}$ is a weight matrix. This projection is task-specific. It can be interpreted as a linear classifier for node classification or regarded as a projection to the space with node similarity measures for link prediction.

\subsection{Metapath Instance Encoders}
\label{sec:metapath-instance-encoders}
To encode each metapath instance in Section~\ref{sec:node_level}, we examine three candidate encoder functions:
\begin{itemize}
    \item \textbf{Mean encoder}. This function takes the element-wise mean of the node vectors along the metapath instance $P\left(v,u\right)$:
    \begin{equation}
        \mathbf{h}_{P\left(v,u\right)}=\operatorname{MEAN}\left(\left\{\mathbf{h}_{t}^{\prime}, \forall t \in P\left(v, u\right)\right\}\right).
    \end{equation}
    \item \textbf{Linear encoder}. This function is an extension to the mean encoder by appending it with a linear transformation:
    \begin{equation}
        \mathbf{h}_{P\left(v,u\right)}=\mathbf{W}_{P} \cdot \operatorname{MEAN}\left(\left\{\mathbf{h}_{t}^{\prime}, \forall t \in P\left(v,u\right)\right\}\right).
    \end{equation}
    \item \textbf{Relational rotation encoder}. We also examine a metapath instance encoder based on relational rotation in complex space, an operation proposed by RotatE~\cite{DBLP:conf/iclr/SunDNT19} for knowledge graph embedding. The mean and linear encoders introduced above treat the metapath instance essentially as a set, and thus ignore the information embedded in the sequential structure of the metapath. Relational rotation provides a way to model this kind of knowledge. Given $P\left(v,u\right)=\left(t_{0},t_{1}, \ldots, t_{n}\right)$ with $t_{0}=u$ and $t_{n}=v$, let $R_{i}$ be the relation between node $t_{i-1}$ and node $t_{i}$, let $\mathbf{r}_{i}$ be the relation vector of $R_{i}$, the relational rotation encoder is formulated as:
    \begin{equation}
    \begin{split}
        &\mathbf{o}_{0} = \mathbf{h}_{t_{0}}^{\prime}=\mathbf{h}_{u}^{\prime}, \\
        &\mathbf{o}_{i} = \mathbf{h}_{t_{i}}^{\prime} + \mathbf{o}_{i-1} \odot \mathbf{r}_{i}, \\
        &\mathbf{h}_{P\left(v,u\right)} = \frac{\mathbf{o}_{n}}{n+1},
    \end{split}
    \end{equation}
    where $\mathbf{h}_{t_{i}}^{\prime}$ and $\mathbf{r}_{i}$ are both complex vectors, $\odot$ is the element-wise product. We can easily interpret a real vector of dimension $d^{\prime}$ as a complex vector of dimension $d^{\prime}/2$ by treating the first half of the vector as the real part, and the second half as the imaginary part.
\end{itemize}

%%% Jenny: Should we illustrate the differences and similarities of these encoders

\subsection{Training}
After applying components introduced in the previous sections, we obtain the final node representations, which can then be used in different downstream tasks. Depending on the characteristics of different tasks and the availability of node labels, we can train MAGNN in two major learning paradigms, i.e., semi-supervised learning and unsupervised learning.

For semi-supervised learning, with the guide of a small fraction of labeled nodes, we can optimize the model weights by minimizing the cross entropy via backpropagation and gradient descent, and thereby learn meaningful node embeddings for heterogeneous graphs. The cross entropy loss for this semi-supervised learning is formulated as:
\begin{equation} \label{eq:semi-supervised-loss}
    \mathcal{L} = - \sum_{v \in \mathcal{V}_{L}} \sum_{c=1}^{C} \mathbf{y}_{v}[c] \cdot \log \mathbf{h}_{v}[c],
\end{equation}
where $\mathcal{V}_{L}$ is the set of nodes that have labels, $C$ is the number of classes, $\mathbf{y}_{v}$ is the one-hot label vector of node $v$, and $\mathbf{h}_{v}$ is the predicted probability vector of node $v$.

For unsupervised learning, without any node labels, we can optimize the model weights by minimizing the following loss function through negative sampling~\cite{NIPS2013_5021}:
\begin{equation} \label{eq:unsupervised-loss}
    \mathcal{L} = -{\sum_{\left(u,v\right) \in \Omega}\log\sigma\left(\mathbf{h}_{u}^{\intercal} \cdot \mathbf{h}_{v} \right)} - {\sum_{\left(u^{\prime},v^{\prime}\right) \in \Omega^{-}}\log\sigma\left(-\mathbf{h}_{u^{\prime}}^{\intercal} \cdot \mathbf{h}_{v^{\prime}}\right)},
\end{equation}
where $\sigma(\cdot)$ is the sigmoid function, $\Omega$ is the set of observed (positive) node pairs, $\Omega^{-}$ is the set of negative node pairs sampled from all unobserved node pairs (the complement of $\Omega$).

\begin{algorithm}[t]
\caption{MAGNN forward propagation.}
\label{algo:MAGNN}

\SetAlgoLined
\KwIn{The heterogeneous graph $\mathcal{G}=\left(\mathcal{V},\mathcal{E}\right)$, \newline
      %node type mapping function $\phi : \mathcal{V} \rightarrow \mathcal{A}$, \newline
      node types $\mathcal{A} = \left\{A_{1}, A_{2}, \ldots, A_{|\mathcal{A}|}\right\}$, \newline
      metapaths $\mathcal{P} = \left\{P_{1}, P_{2}, \ldots P_{|\mathcal{P}|}\right\}$, \newline
      %edge type mapping function $\psi : \mathcal{E} \rightarrow \mathcal{R}$, \newline
      node features $\left\{\mathbf{x}_{v}, \forall v \in \mathcal{V}\right\}$, \newline
      the number of attention heads $K$, \newline
      the number of layers $L$
      }
\KwOut{The node embeddings $\left\{\mathbf{z}_{v}, \forall v \in \mathcal{V}\right\}$}

\For{node type $A \in \mathcal{A}$}{
  Node content transformation $\mathbf{h}_{v}^{0} \leftarrow \mathbf{W}_{A} \cdot \mathbf{x}_{v}, \forall v \in \mathcal{V}_{A}$\;
}
\For{$l = 1 \ldots L$}{
  \For{node type $A \in \mathcal{A}$}{
    \For{metapath $P \in \mathcal{P}_{A}$}{
      %\tcc{intra-metapath aggregation}
      %\textcolor{blue}{
      \For{$v \in \mathcal{V}_{A}$}{
        Calculate $\mathbf{h}_{P\left(v,u\right)}^{l}$ for all $u \in \mathcal{N}_{v}^{P}$ using the metapath instance encoder function\;
        Combine extracted metapath instances $\left[\mathbf{h}_{v}^{P}\right]^{l} \leftarrow \overset{K}{\underset{k=1}{\|}} \sigma\left(\sum_{u \in \mathcal{N}_{v}^{P}} \left[\alpha_{v u}^{P}\right]_{k} \cdot \mathbf{h}_{P\left(v,u\right)}^{l}\right)$\;
      }
      %}
    }
    %\tcc{inter-metapath aggregation}
    %\textcolor{red}{
    Calculate the weight $\beta_{P}$ for each metapath $P \in \mathcal{P}_{A}$\;
    Fuse the embeddings from different metapaths $\left[\mathbf{h}_{v}^{\mathcal{P}_{A}}\right]^{l} \leftarrow \sum_{P \in \mathcal{P}_{A}} \beta_{P} \cdot \left[\mathbf{h}_{v}^{P}\right]^{l}, \forall v \in \mathcal{V}_{A}$\;
    %}
  }
  Layer output projection $\mathbf{h}_{v}^{l} = \sigma\left(\mathbf{W}_{o}^{l} \cdot \left[\mathbf{h}_{v}^{\mathcal{P}_{A}}\right]^{l}\right), \forall v \in \mathcal{V}_{A}, \forall A \in \mathcal{A}$\;
}
$\mathbf{z}_{v} \leftarrow \mathbf{h}_{v}^{L}, \forall v \in \mathcal{V}$\;

\end{algorithm}

\begin{figure*}[ht!]
     \centering
     \begin{subfigure}[b]{0.3\textwidth}
         \centering
         \includegraphics[width=\textwidth]{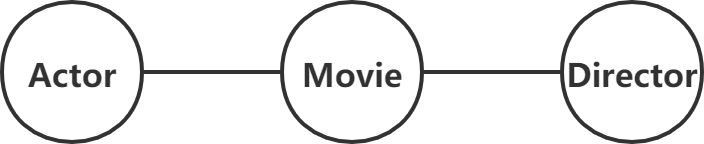}
         \caption{IMDb}
     \end{subfigure}
     \hfill
     \begin{subfigure}[b]{0.3\textwidth}
         \centering
         \includegraphics[width=\textwidth]{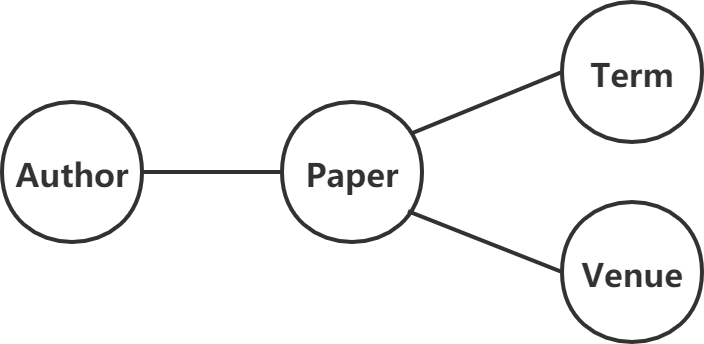}
         \caption{DBLP}
     \end{subfigure}
     \hfill
     \begin{subfigure}[b]{0.3\textwidth}
         \centering
         \includegraphics[width=\textwidth]{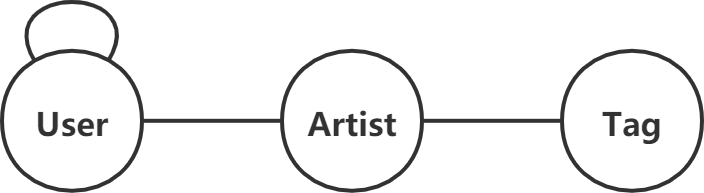}
         \caption{Last.fm}
     \end{subfigure}
      \caption{Network schemas of the three heterogeneous graph datasets used in this paper.}
      \label{fig:network_schemas}
\end{figure*}

\section{Experiments}

In this section, we present experiments to demonstrate the efficacy of MAGNN for heterogeneous graph embedding. The experiments aim to address the following research questions:
\begin{itemize}
    \item RQ1. How does MAGNN perform in classifying nodes? % Node classification
    \item RQ2. How does MAGNN perform in clustering nodes? % Node clustering
    \item RQ3. How does MAGNN perform in predicting plausible links between node pairs? % Link Prediction
    \item RQ4. What is the impact of the three major components of MAGNN described in the previous section? % Ablation studies
    \item RQ5. How do we understand the representation capability of different graph embedding methods? % Visualization
\end{itemize}

\subsection{Datasets}

We adopt three widely used heterogeneous graph datasets from different domains to evaluate the performance of MAGNN as compared to state-of-the-art baselines. Specifically, the IMDb and DBLP datasets are used in the experiments for node classification and node clustering. The Last.fm dataset is used in the experiments for link prediction. Simple statistics of the three datasets are summarized in Table~\ref{tab:dataset}, and network schemas are illustrated in Figure~\ref{fig:network_schemas}. We assign one-hot id vectors to nodes with no attributes as their dummy input features.
\begin{itemize}
    \item \textbf{IMDb}\footnote{\url{https://www.imdb.com/}} is an online database about movies and television programs, including information such as cast, production crew, and plot summaries. We use a subset of IMDb scraped from online, containing 4278 movies, 2081 directors, and 5257 actors after data preprocessing. Movies are labeled as one of three classes (\textit{Action}, \textit{Comedy}, and \textit{Drama}) based on their genre information. Each movie is also described by a bag-of-words representation of its plot keywords. For semi-supervised learning models, the movie nodes are divided into training, validation, and testing sets of 400 (9.35\%), 400 (9.35\%), and 3478 (81.30\%) nodes, respectively.
    \item \textbf{DBLP}\footnote{\url{https://dblp.uni-trier.de/}} is a computer science bibliography website. We adopt a subset of DBLP extracted by \cite{Gao:2009:GCM:2984093.2984159,Ji:2010:GRT:1888258.1888302}, containing 4057 authors, 14328 papers, 7723 terms, and 20 publication venues after data preprocessing. The authors are divided into four research areas (\textit{Database}, \textit{Data Mining}, \textit{Artificial Intelligence}, and \textit{Information Retrieval}). Each author is described by a bag-of-words representation of their paper keywords. For semi-supervised learning models, the author nodes are divided into training, validation, and testing sets of 400 (9.86\%), 400 (9.86\%), and 3257 (80.28\%) nodes, respectively.
    \item \textbf{Last.fm}\footnote{\url{https://www.last.fm/}} is a music website keeping track of users' listening information from various sources. We adopt a dataset released by HetRec~2011~\cite{Cantador:RecSys2011}, consisting of 1892 users, 17632 artists, and 1088 artist tags after data preprocessing. This dataset is used for the link prediction task, and no label or feature is included in this dataset. For semi-supervised learning models, the user-artist pairs are divided into training, validation, and testing sets of 64984 (70\%), 9283 (10\%), and 18567 (20\%) pairs, respectively.
\end{itemize}

\begin{table}[t]
  \caption{Statistics of datasets.}
  \label{tab:dataset}
  \begin{tabular}{c||c|c|c}
    \toprule
    Dataset & Node & Edge & Metapath\\
    \midrule
    IMDb & \specialcell{\# movie (M): 4,278 \\ \# director (D): 2,081 \\ \# actor (A): 5,257} & \specialcell{\# M-D: 4,278 \\ \# M-A: 12,828} & \specialcell{MDM \\ MAM \\ DMD \\ DMAMD \\ AMA \\ AMDMA}\\
    \midrule
    DBLP & \specialcell{\# author (A): 4,057 \\ \# paper (P): 14,328 \\ \# term (T): 7,723 \\ \# venue (V): 20} & \specialcell{\# A-P: 19,645 \\ \# P-T: 85,810 \\ \# P-V: 14,328} & \specialcell{APA \\ APTPA \\ APVPA}\\
    \midrule
    Last.fm & \specialcell{\# user (U): 1,892 \\ \# artist (A): 17,632 \\ \# tag (T): 1,088} & \specialcell{\# U-U: 12,717 \\ \# U-A: 92,834 \\ \# A-T: 23,253} & \specialcell{UU \\ UAU \\ UATAU \\ AUA \\ AUUA \\ ATA}\\
%    Last.fm & \specialcell{\# user (U): 1892 \\ \# artist (A): 17632 \\ \# tag (T) (reduced): 1088 \\ \# tag (T) (original): 11945} & \specialcell{\# U-U: 12717 \\ \# U-A: 92834 \\ \# A-T (reduced): 23253 \\ \# A-T (complete): 108437} & \specialcell{UU \\ UAU \\ UATAU \\ AUA \\ AUUA \\ ATA}\\
    \bottomrule
  \end{tabular}
\end{table}

\subsection{Baselines}

\begin{table*}[ht!]
\caption{Experiment results (\%) on the IMDb and DBLP datasets for the node classification task. }
\label{tab:node_class}
\begin{tabular}{|c|c|c|c|c|c|c|c|c|c|c|c|}
\hline
\multirow{2}{*}{Dataset} & \multirow{2}{*}{Metrics}  & \multirow{2}{*}{Train \%} & \multicolumn{5}{c|}{Unsupervised}                          & \multicolumn{4}{c|}{Semi-supervised}   \\ \cline{4-12}
                         &                           &                           & LINE  & node2vec & ESim  & metapath2vec & HERec & GCN   & GAT   & HAN   & MAGNN          \\ \hline
\multirow{8}{*}{IMDb} & \multirow{4}{*}{Macro-F1} & 20\%       & 44.04 & 49.00    & 48.37 & 46.05        & 45.61 & 52.73 & 53.64 & 56.19 & \textbf{59.35} \\ \cline{3-12}
                      &                           & 40\%       & 45.45 & 50.63    & 50.09 & 47.57        & 46.80 & 53.67 & 55.50 & 56.15 & \textbf{60.27} \\ \cline{3-12}
                      &                           & 60\%       & 47.09 & 51.65    & 51.45 & 48.17        & 46.84 & 54.24 & 56.46 & 57.29 & \textbf{60.66} \\ \cline{3-12}
                      &                           & 80\%       & 47.49 & 51.49    & 51.37 & 49.99        & 47.73 & 54.77 & 57.43 & 58.51 & \textbf{61.44} \\ \cline{2-12}
                      & \multirow{4}{*}{Micro-F1} & 20\%       & 45.21 & 49.94    & 49.32 & 47.22        & 46.23 & 52.80 & 53.64 & 56.32 & \textbf{59.60} \\ \cline{3-12}
                      &                           & 40\%       & 46.92 & 51.77    & 51.21 & 48.17        & 47.89 & 53.76 & 55.56 & 57.32 & \textbf{60.50} \\ \cline{3-12}
                      &                           & 60\%       & 48.35 & 52.79    & 52.53 & 49.87        & 48.19 & 54.23 & 56.47 & 58.42 & \textbf{60.88} \\ \cline{3-12}
                      &                           & 80\%       & 48.98 & 52.72    & 52.54 & 50.50        & 49.11 & 54.63 & 57.40 & 59.24 & \textbf{61.53} \\ \hline
\multirow{8}{*}{DBLP} & \multirow{4}{*}{Macro-F1} & 20\%       & 87.16 & 86.70    & 90.68 & 88.47        & 90.82 & 88.00 & 91.05 & 91.69 & \textbf{93.13} \\ \cline{3-12}
                      &                           & 40\%       & 88.85 & 88.07    & 91.61 & 89.91        & 91.44 & 89.00 & 91.24 & 91.96 & \textbf{93.23} \\ \cline{3-12}
                      &                           & 60\%       & 88.93 & 88.69    & 91.84 & 90.50        & 92.08 & 89.43 & 91.42 & 92.14 & \textbf{93.57} \\ \cline{3-12}
                      &                           & 80\%       & 89.51 & 88.93    & 92.27 & 90.86        & 92.25 & 89.98 & 91.73 & 92.50 & \textbf{94.10} \\ \cline{2-12}
                      & \multirow{4}{*}{Micro-F1} & 20\%       & 87.68 & 87.21    & 91.21 & 89.02        & 91.49 & 88.51 & 91.61 & 92.33 & \textbf{93.61} \\ \cline{3-12}
                      &                           & 40\%       & 89.25 & 88.51    & 92.05 & 90.36        & 92.05 & 89.22 & 91.77 & 92.57 & \textbf{93.68} \\ \cline{3-12}
                      &                           & 60\%       & 89.34 & 89.09    & 92.28 & 90.94        & 92.66 & 89.57 & 91.97 & 92.72 & \textbf{93.99} \\ \cline{3-12}
                      &                           & 80\%       & 89.96 & 89.37    & 92.68 & 91.31        & 92.78 & 90.33 & 92.24 & 93.23 & \textbf{94.47} \\ \hline

\end{tabular}
\end{table*}

We compare MAGNN against different kinds of graph embedding models, including traditional (as opposed to GNNs) homogeneous graph embedding models, traditional heterogeneous graph embedding models, GNNs for homogeneous graphs, and GNNs for heterogeneous graphs. We denote them as \emph{traditional homogeneous models}, \emph{traditional heterogeneous models}, \emph{homogeneous GNNs}, and \emph{heterogeneous GNNs}, respectively. The list of baseline models is shown as follows.
\begin{itemize}
%    \item \textbf{DeepWalk}~\cite{Perozzi:2014:DOL:2623330.2623732} is a \emph{homogeneous} model based on random walks and a skip-gram model. We apply DeepWalk to the heterogeneous graphs by ignoring the heterogeneity and dropping node content features.
    \item \textbf{LINE}~\cite{Tang:2015:LLI:2736277.2741093} is a \emph{traditional homogeneous model} exploiting the first-order and second-order proximity between nodes. We apply it to the heterogeneous graphs by ignoring the heterogeneity of graph structure and dropping all node content features. The LINE variant using second-order proximity is applied in our experiments.
    \item \textbf{node2vec}~\cite{Grover:2016:NSF:2939672.2939754} is a \emph{traditional homogeneous model} serving as a generalized version of DeepWalk~\cite{Perozzi:2014:DOL:2623330.2623732}. We apply it to the heterogeneous graphs in the same way as LINE.
    \item \textbf{ESim}~\cite{DBLP:journals/corr/ShangQLKHP16} is a \emph{traditional heterogeneous model} that learns node embeddings from sampled metapath instances. ESim requires a predefined weight for each metapath. Here we assign equal weights to all metapaths because searching for the optimal weights of metapaths is difficult, and does not provide a significant performance gain over equal weights according to the authors’ experiments.
    \item \textbf{metapath2vec}~\cite{Dong:2017:MSR:3097983.3098036} is a \emph{traditional heterogeneous model} that generates node embeddings by feeding metapath-guided random walks to a skip-gram model. This model relies on a single user-specified metapath, so we test on all metapaths separately and report the one with the best results. We use the metapath2vec++ model variant in our experiments.
    \item \textbf{HERec}~\cite{8355676} is a \emph{traditional heterogeneous model} that learns node embeddings by applying DeepWalk to the metapath-based homogeneous graphs converted from the original heterogeneous graph. This model comes with an embedding fusion algorithm designed for rating prediction, which can be adapted to link prediction. For node classification/clustering, we select and report the metapath with the best performance.
    \item \textbf{GCN}~\cite{DBLP:conf/iclr/KipfW17} is a \emph{homogeneous GNN}. This model performs convolutional operations in the graph Fourier domain. Here we test GCN on metapath-based homogeneous graphs and report the results from the best metapath.
    \item \textbf{GAT}~\cite{DBLP:conf/iclr/VelickovicCCRLB18} is a \emph{homogeneous GNN}. This model performs convolutional operations in the graph spatial domain with the attention mechanism incorporated. Similarly, here we test GAT on metapath-based homogeneous graphs and report the results from the best metapath.
    \item \textbf{GATNE}~\cite{cen2019representation} is a \emph{heterogeneous GNN}. It generates a node's representation from the base embedding and the edge embeddings, with a focus on the link prediction task. Here we report the results from the best-performing GATNE variant.
    \item \textbf{HAN}~\cite{Wang:2019:HGA:3308558.3313562} is a \emph{heterogeneous GNN}. It learns metapath-specific node embeddings from different metapath-based homogeneous graphs, and leverages the attention mechanism to combine them into one vector representation for each node.
    %\item \textbf{MAGNN} is our metapath aggregated graph neural network with the relational rotation metapath instance encoder.
\end{itemize}

For traditional models, including LINE, node2vec, ESim, metapath2vec, and HERec, we set the window size to 5, walk length to 100, walks per node to 40, and number of negative samples to 5, if applicable.
For GNNs, including GCN, GAT, HAN, and our proposed MAGNN, we set the dropout rate to 0.5; we use the same splits of training, validation, and testing sets; we employ the Adam optimizer with the learning rate set to 0.005 and the weight decay (L2 penalty) set to 0.001; we train the GNNs for 100 epochs and apply early stopping with a patience of 30. For node classification and node clustering, the GNNs are trained in a semi-supervised fashion with a small fraction of nodes labeled as guidance.
For GAT, HAN, and MAGNN, we set the number of attention heads to 8.
For HAN and MAGNN, we set the dimension of the attention vector in inter-metapath aggregation to 128.
For a fair comparison, we set the embedding dimension of all the models mentioned above to 64.

\begin{table*}[ht!]
\caption{Experiment results (\%) on the IMDb and DBLP datasets for the node clustering task. }
\label{tab:node_clust}
\begin{tabular}{|c|c|c|c|c|c|c|c|c|c|c|}
\hline
\multirow{2}{*}{Dataset} & \multirow{2}{*}{Metrics} & \multicolumn{5}{c|}{Unsupervised}                          & \multicolumn{4}{c|}{Semi-supervised}   \\ \cline{3-11}
                         &                          & LINE  & node2vec & ESim  & metapath2vec & HERec & GCN   & GAT   & HAN   & MAGNN          \\ \hline
\multirow{2}{*}{IMDb} & NMI     & 1.13  & 5.22     & 1.07  & 0.89         & 0.39  & 7.46  & 7.84  & 10.79 & \textbf{15.58} \\ \cline{2-11}
                      & ARI     & 1.20  & 6.02     & 1.01  & 0.22         & 0.11  & 7.69  & 8.87  & 11.11 & \textbf{16.74} \\ \hline
\multirow{2}{*}{DBLP} & NMI     & 71.02 & 77.01    & 68.33 & 74.18        & 69.03 & 73.45 & 70.73 & 77.49 & \textbf{80.81} \\ \cline{2-11}
                      & ARI     & 76.52 & 81.37    & 72.22 & 78.11        & 72.45 & 77.50 & 76.04 & 82.95 & \textbf{85.54} \\ \hline
\end{tabular}
\end{table*}

\begin{table*}[ht!]
\caption{Experiment results (\%) on the Last.fm dataset for the link prediction task. }
\label{tab:link_pred}
\begin{tabular}{|c|c|c|c|c|c|c|c|c|c|c|c|}
\hline
Dataset                  & Metrics & LINE  & node2vec & ESim  & metapath2vec & HERec & GCN   & GAT   & GATNE & HAN   & MAGNN          \\ \hline
\multirow{2}{*}{Last.fm} & AUC     & 85.76 & 67.14    & 82.00 & 92.20        & 91.52 & 90.97 & 92.36 & 89.21 & 93.40 & \textbf{98.91} \\ \cline{2-12}
                         & AP      & 88.07 & 64.11    & 82.19 & 90.11        & 89.47 & 91.65 & 91.55 & 88.86 & 92.44 & \textbf{98.93} \\ \hline
\end{tabular}
\end{table*}

\subsection{Node Classification (RQ1)}
\label{sec:node_class}
We conduct experiments on the IMDb and DBLP datasets to compare the performance of different models on the node classification task.
We feed the embeddings of labeled nodes (movies in IMDb and authors in DBLP) generated by each learning model to a linear support vector machine (SVM) classifier with varying training proportions. Note that for a fair comparison, only the nodes in the testing set are fed to the linear SVM, because semi-supervised models have already ``seen'' the nodes in the training and validation sets, as shown in Equation~\ref{eq:semi-supervised-loss}.
Hence, the training and testing proportions of the linear SVM here only concern the testing set (i.e., 3478 nodes for IMDb and 3257 nodes for DBLP).
Again, the train/test splits for the linear SVM are also the same across embedding models. Similar strategies are also applied to the experiments of node clustering and link prediction. We report the average \emph{Macro-F1} and \emph{Micro-F1} of 10 runs of each embedding model in Table~\ref{tab:node_class}.

As shown in the table, MAGNN performs consistently better than other baselines across different training proportions and datasets.
On IMDb, it is interesting to see that node2vec performs better than traditional heterogeneous models. That said, GNNs, especially heterogeneous GNNs, obtain even better results, demonstrating that the GNN architecture, which judiciously utilizes the heterogeneous node features, helps improve the embedding performance. The performance gain obtained by MAGNN over the best baseline (HAN) is around 4-7\%, which indicates that metapath instances contain richer information than metapath-based neighbors.
On DBLP, the node classification task is trivial, as evident from the high scores of all models. Even so, MAGNN still outperforms the strongest baseline by 1-2\%.

\subsection{Node Clustering (RQ2)}
We conduct experiments on the IMDb and DBLP datasets to compare the performance of different models on the node clustering task. We feed the embeddings of labeled nodes (movies in IMDb and authors in DBLP) generated by each learning model to the K-Means algorithm. The number of clusters in K-Means is set to the number of classes for each dataset, i.e., 3 for IMDb and 4 for DBLP. We employ the \emph{normalized mutual information} (NMI) and \emph{adjusted Rand index} (ARI) as the evaluation metrics.
Since the clustering result of the K-Means algorithm is highly dependent on the initialization of the centroids, we repeat K-Means 10 times for each run of the embedding model, and each embedding model is tested for 10 runs. We report the averaged results in Table~\ref{tab:node_clust}.

From Table~\ref{tab:node_clust}, we can see that MAGNN is consistently superior to all other baselines in node clustering.
Note that all models have much poorer performance on IMDb than on DBLP. This is presumably because of the dirty labels of movies in IMDb: every movie node in the original IMDb dataset has multiple genres, and we only choose the very first one as its class label. We can see that the traditional heterogeneous models do not have many advantages over the traditional homogeneous models in node clustering. Node2vec is expected to perform strongly in the node clustering task because, being a random-walk-based approach, it forces nodes that are close in the graph also to be close in the embedding space \cite{DBLP:conf/icml/YouYL19}, and thereby encodes node positional information. This property implicitly facilitates the K-Means algorithm as it clusters nodes based on the Euclidean distances between embeddings. Despite this, the heterogeneity-aware GNNs (i.e., HAN and MAGNN) still rank the first in node clustering on both datasets.

\subsection{Link Prediction (RQ3)}

We also conduct experiments on the Last.fm dataset to evaluate the performance of MAGNN and other baselines in the link prediction task.
For the GNNs, we treat the connected user-artist pair as positive node pairs, and consider all unconnected user-artist links as negative node pairs. We add the same number of randomly sampled negative node pairs to the validation and testing sets. During the GNNs' training, negative node pairs are also uniformly sampled on the fly.
The GNNs are then optimized by minimizing the objective function described in Equation~\ref{eq:unsupervised-loss}.

Given the user embedding $\mathbf{h}_{u}$ and the artist embedding $\mathbf{h}_{a}$ generated by the trained model, we calculate the probability that $u$ and $v$ link together as follows:
\begin{equation}
    p_{u a} = \sigma\left(\mathbf{h}_{u}^{\intercal} \cdot \mathbf{h}_{a}\right),
\end{equation}
where $\sigma(\cdot)$ is the sigmoid function.
The embedding models for link prediction are evaluated by the \emph{area under the ROC curve} (AUC) and \emph{average precision} (AP) scores. We report the averaged results of 10 runs of each embedding model in Table~\ref{tab:link_pred}.

From Table~\ref{tab:link_pred}, MAGNN outperforms other baseline models by a large margin. The strongest traditional model here is metapath2vec, which learns from node sequences generated from random walks guided by a single metapath. MAGNN achieves better scores than metapath2vec, showing that considering a single metapath is suboptimal. Among GNN baselines, HAN obtains the best results because it is heterogeneity-aware and combines multiple metapaths. Our MAGNN achieves a relative improvement of around 6\% over HAN. This result supports our claim that the metapath contexts of nodes are critical to the node embeddings.

%In real-world graphs, different nodes usually have different distributions of metapaths. Hence, to handle this imbalance, it is significant to include importance computations of different types of pre-defined metapaths for each node in our model.

\begin{table*}[ht!]
\caption{Quantitative results (\%) for ablation study.}
\label{tab:ablation}
\begin{tabular}{|l|c|c|c|c|c|c|c|c|c|c|}
\hline
\multicolumn{1}{|c|}{\multirow{2}{*}{Variant}} & \multicolumn{4}{c|}{IMDb}           & \multicolumn{4}{c|}{DBLP}           & \multicolumn{2}{c|}{Last.fm} \\ \cline{2-11}
\multicolumn{1}{|c|}{}                         & Macro-F1 & Micro-F1 & NMI   & ARI   & Macro-F1 & Micro-F1 & NMI   & ARI   & AUC           & AP           \\ \hline
$\mathrm{MAGNN}_{\mathrm{feat}}$                                    & 48.87    & 50.36    & 5.82  & 5.30  & 92.80    & 93.32    & 77.17 & 82.15 & N/A           & N/A          \\ \hline
$\mathrm{MAGNN}_{\mathrm{nb}}$                                      & 58.45    & 58.84    & 12.87 & 11.98 & 92.61    & 93.15    & 77.64 & 82.60 & 93.68         & 92.95        \\ \hline
$\mathrm{MAGNN}_{\mathrm{sm}}$                                      & 56.77    & 56.64    & 11.90 & 11.84 & 93.19    & 93.69    & 79.48 & 84.39 & 92.54         & 91.52        \\ \hline
$\mathrm{MAGNN}_{\mathrm{avg}}$                                     & 59.66    & 59.78    & 13.64 & 15.27 & 93.13    & 93.44    & 79.31 & 84.30 & 98.63         & 98.57        \\ \hline
$\mathrm{MAGNN}_{\mathrm{linear}}$                                  & 57.80    & 57.96    & 9.80  & 8.49  & 93.21    & 93.52    & 78.95 & 83.89 & 98.56         & 98.48        \\ \hline
$\mathrm{MAGNN}_{\mathrm{rot}}$                                     & \textbf{60.43}    & \textbf{60.63}    & \textbf{15.58} & \textbf{16.74} & \textbf{93.51}    & \textbf{93.94}    & \textbf{80.81} & \textbf{85.54} & \textbf{98.91}         & \textbf{98.93}        \\ \hline
\end{tabular}
\end{table*}

\subsection{Ablation Study (RQ4)}

To validate the effectiveness of each component of our model, we further conduct experiments on different MAGNN variants.
Here we report the results obtained from the three datasets on all three tasks in Table~\ref{tab:ablation}. Note that every presented score of the node classification task (i.e., Macro-F1 and Micro-F1) is an average of the scores in different training proportions (explained in Section \ref{sec:node_class}).
Here $\mathrm{MAGNN}_{\mathrm{rot}}$ is our proposed model using the relational rotation encoder, i.e., the one used to compete with other baselines in Table~\ref{tab:node_class}, \ref{tab:node_clust}, and \ref{tab:link_pred}.
Let $\mathrm{MAGNN}_{\mathrm{rot}}$ be the reference model, $\mathrm{MAGNN}_{\mathrm{feat}}$ is the equivalent model without utilizing node content features; $\mathrm{MAGNN}_{\mathrm{nb}}$ considers only the metapath-based neighbors; $\mathrm{MAGNN}_{\mathrm{sm}}$ considers the single best metapath; $\mathrm{MAGNN}_{\mathrm{avg}}$ switches to using the mean metapath instance encoder; $\mathrm{MAGNN}_{\mathrm{linear}}$ switches to using the linear metapath instance encoder. Except for the above-mentioned differences, all other settings are the same for these MAGNN variants. Note that $\mathrm{MAGNN}_{\mathrm{feat}}$ on Last.fm is equivalent to $\mathrm{MAGNN}_{\mathrm{rot}}$ because this dataset does not contain node attributes.

As can be seen, by utilizing the node content features, $\mathrm{MAGNN}_{\mathrm{rot}}$ obtains a significant performance improvement over $\mathrm{MAGNN}_{\mathrm{feat}}$, which shows the necessity of applying node content transformation to incorporate node features.
Comparing $\mathrm{MAGNN}_{\mathrm{nb}}$ with $\mathrm{MAGNN}_{\mathrm{avg}}$, $\mathrm{MAGNN}_{\mathrm{linear}}$, and $\mathrm{MAGNN}_{\mathrm{rot}}$, we see that aggregating metapath instances rather than metapath-based neighbors brings about a boost in performance, which validates the efficacy of intra-metapath aggregation.
Next, the difference between the results of $\mathrm{MAGNN}_{\mathrm{sm}}$ and $\mathrm{MAGNN}_{\mathrm{rot}}$ reveals that the model performance is improved considerably by combining multiple metapaths in inter-metapath aggregation.
Finally, the results of $\mathrm{MAGNN}_{\mathrm{avg}}$, $\mathrm{MAGNN}_{\mathrm{linear}}$, and $\mathrm{MAGNN}_{\mathrm{rot}}$ suggest that the relational rotation encoder does help to improve MAGNN by a small margin.
It is interesting to see that $\mathrm{MAGNN}_{\mathrm{linear}}$ performs worse than $\mathrm{MAGNN}_{\mathrm{avg}}$. Nonetheless, all three MAGNN variants using different encoders still consistently outperform the best baseline, HAN.

\subsection{Visualization (RQ5)}

\begin{figure}[t]
     \centering
     \begin{subfigure}[b]{0.49\linewidth}
         \centering
         \includegraphics[width=\textwidth]{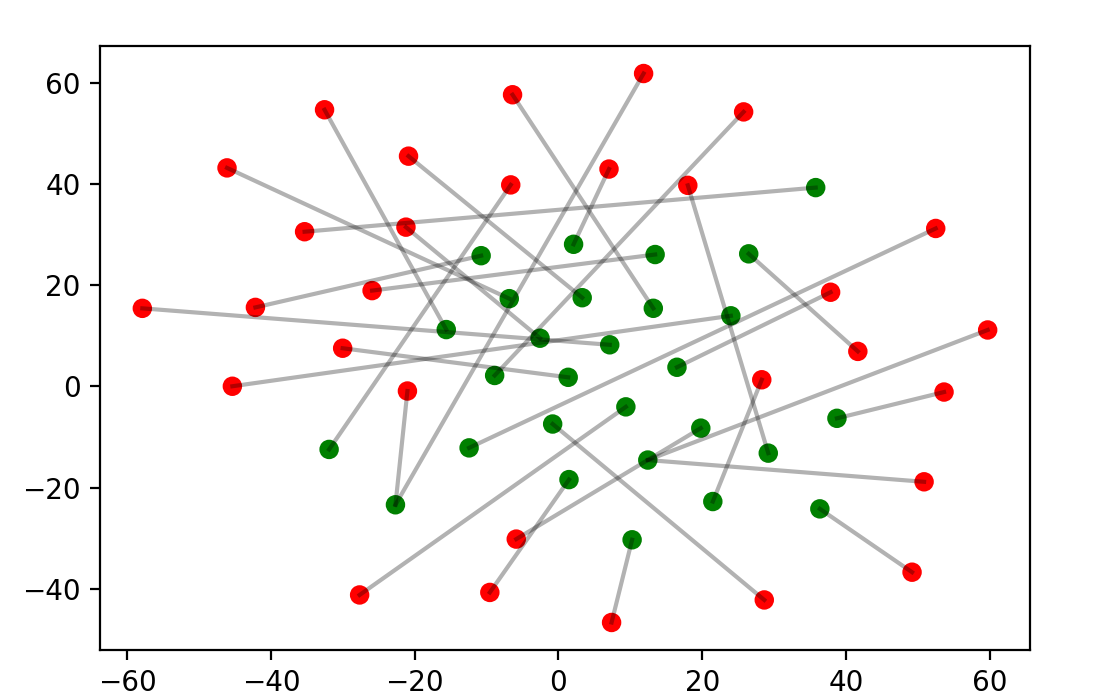}
         \caption{LINE}
     \end{subfigure}
     \hfill
     \begin{subfigure}[b]{0.49\linewidth}
         \centering
         \includegraphics[width=\textwidth]{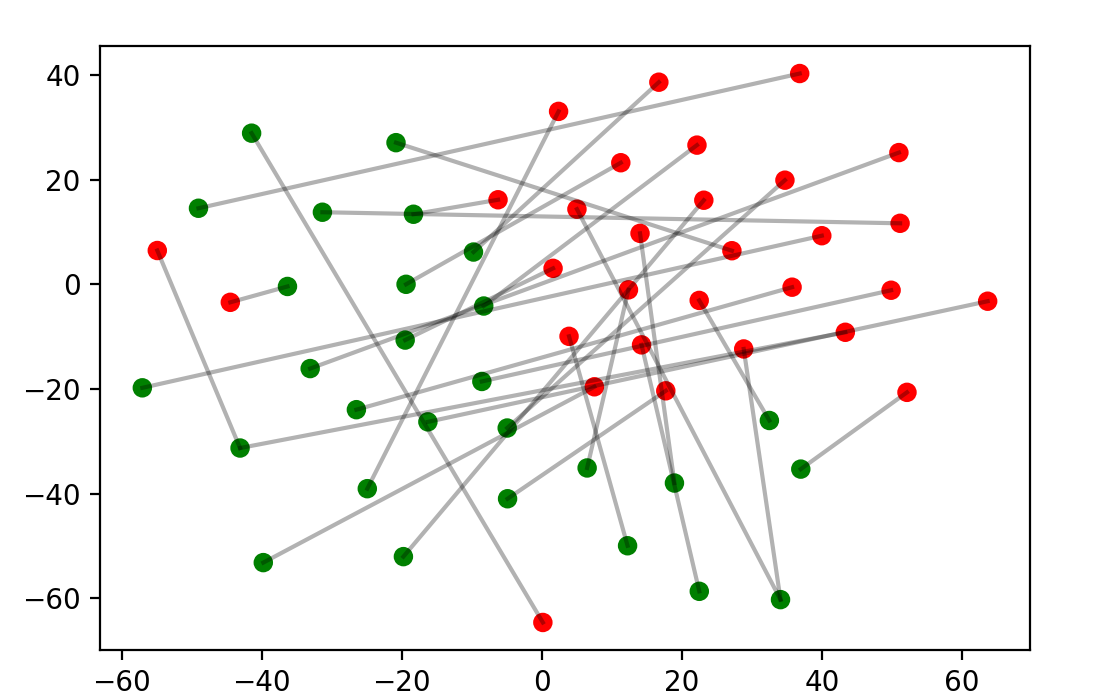}
         \caption{ESim}
     \end{subfigure}
     \hfill
     \begin{subfigure}[b]{0.49\linewidth}
         \centering
         \includegraphics[width=\textwidth]{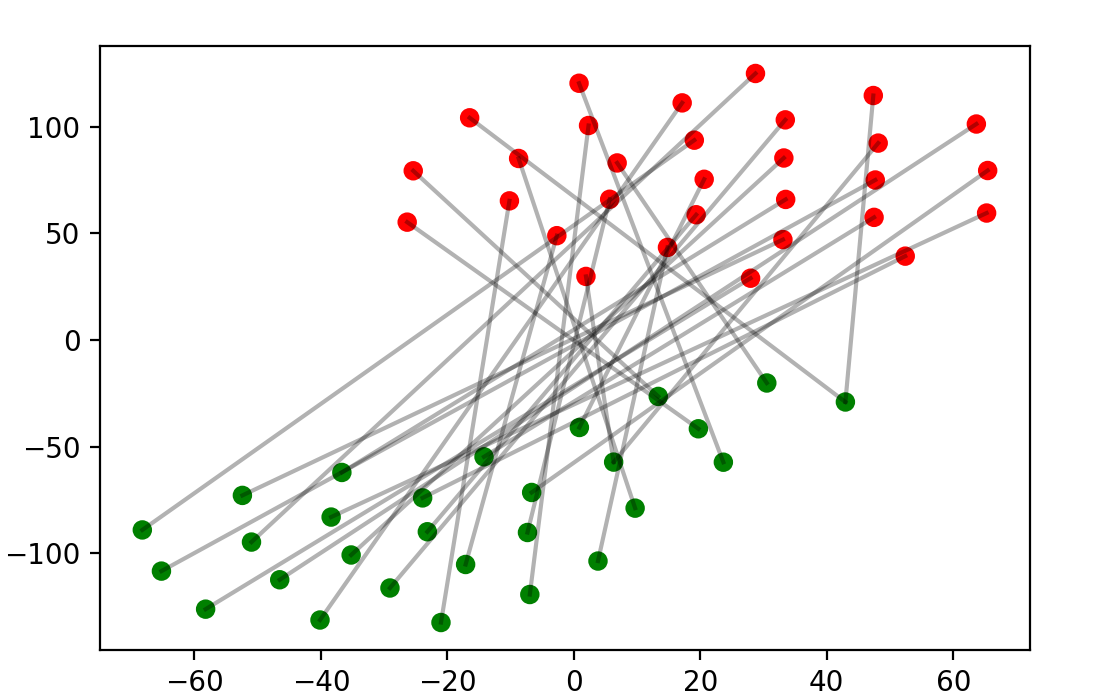}
         \caption{GCN}
     \end{subfigure}
     \hfill
     \begin{subfigure}[b]{0.49\linewidth}
         \centering
         \includegraphics[width=\textwidth]{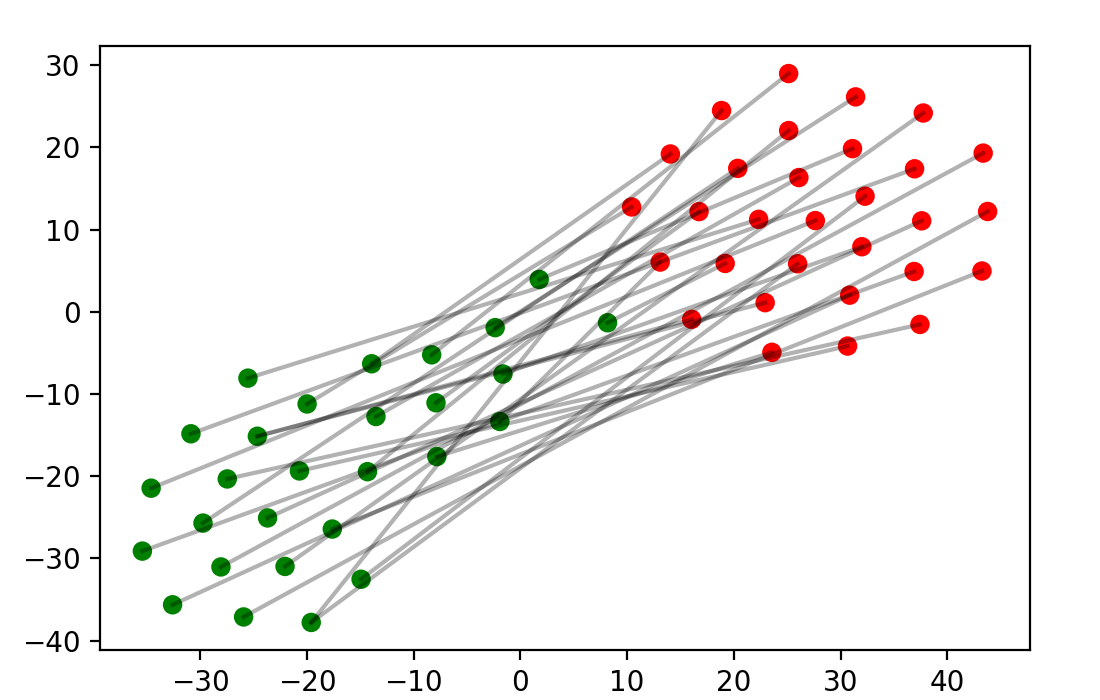}
         \caption{MAGNN}
     \end{subfigure}
      \caption{Embedding visualization of node pairs in Last.fm.}
      \label{fig:visualization}
\end{figure}

In addition to the quantitative evaluations of embedding models, we also visualize node embeddings to conduct a qualitative assessment of the embedding results. We randomly select 30 user-artist pairs from the positive testing set of the Last.fm dataset, and then project the embeddings of these nodes into a 2-dimensional space using t-SNE. Here we illustrate the visualization results of LINE, ESim, GCN, and MAGNN in Figure~\ref{fig:visualization}, where red points and green points indicate users and artists, respectively.

Based on this visualization, one can quickly tell the differences among graph embedding models in terms of their learning ability towards heterogeneous graphs.
As a traditional homogeneous graph embedding model, LINE cannot effectively divide user nodes and artist nodes into two different groups.
In contrast, ESim, a traditional heterogeneous model, can roughly partition the two types of nodes.
Thanks to the powerful GNN architecture and by choosing appropriate metapaths, a homogeneous GNN such as GCN can isolate different types of nodes and encode the correlation information of the user-artist pairs into the node embeddings.
From Figure~\ref{fig:visualization}, we can see that our proposed MAGNN obtains the best embedding results, with two well-separated user and artist groups, and an aligned correlation of user-artist pairs.

\section{Conclusion}

In this paper, we propose a novel metapath aggregated graph neural network (MAGNN) to address the three characteristic limitations of existing heterogeneous graph embedding methods, namely (1) dropping node content features, (2) discarding intermediate nodes along metapaths, and (3) considering only a single metapath.
To be specific, MAGNN applies three building block components: (1) node content transformation, (2) intra-metapath aggregation, and (3) inter-metapath aggregation to deal with each of the limitations, respectively.
Additionally, we define the notion of metapath instance encoders, which are used to extract the structural and semantic information ingrained in metapath instances.
We propose several candidate encoder functions, including one inspired by the RotatE knowledge graph embedding model~\cite{DBLP:conf/iclr/SunDNT19}.
In experiments, MAGNN achieves state-of-the-art results on three real-world datasets in the node classification, node clustering, and link prediction tasks.
Ablation studies also demonstrate the efficacy of the three major components of MAGNN in boosting embedding performance.
We plan to adapt this heterogeneous graph embedding framework to the rating prediction (recommendation) task with the user-item data assisted by the heterogeneous knowledge graph~\cite{Wang:2019:KGC:3308558.3313417}.

\begin{acks}
The work described in this paper was partially supported by the Research Grants Council of the Hong Kong Special Administrative Region, China (CUHK 2300174 (Collaborative Research Fund, No. C5026-18GF) and CUHK 3133238 (Research Sustainability of Major RGC Funding Schemes)).
\end{acks}

\bibliographystyle{ACM-Reference-Format}
\bibliography{reference}

%\appendix



\end{document}